\documentclass[runningheads]{llncs}
\usepackage{subcaption}
\usepackage{graphicx}
\usepackage{xcolor}
\usepackage{soul}
\let\oldmaketitle\maketitle
\renewcommand{\maketitle}{\oldmaketitle\setcounter{footnote}{0}}

\begin{document}

\title{Spot the fake lungs: Generating Synthetic Medical Images using Neural Diffusion Models\thanks{This preprint has not undergone peer review (when applicable) or any post-submission improvements or
corrections. The Version of Record of this contribution is published in [insert volume title],
and is available online at https://doi.org/doi to be updated
. Upon publishing, final version will be available from Springer Nature Publishers (link to be updated).}}
%
\titlerunning{Neural diffusion models for medical image synthesis}
%
\author{Hazrat Ali\inst{1} \and
Shafaq Murad\inst{2} \and
Zubair Shah\inst{1} 
}

\authorrunning{H. Ali et al.}
%
\institute{College of Science and Engineering, Hamad Bin Khalifa University, Qatar Foundation, Doha, Qatar
\and
Manchester University NHS Foundation Trust, Manchester Royal Infirmary, Oxford Road, Manchester, M13 9PL, United Kingdom\\
\email{haali2@hbku.edu.qa, shafaq.murad@mft.nhs.uk, zshah@hbku.edu.qa}
}
\maketitle              
\begin{abstract}
Generative models are becoming popular for the synthesis of medical images. Recently, neural diffusion models have demonstrated the potential to generate photo-realistic images of objects. However, their potential to generate medical images is not explored yet.  
In this work, we explore the possibilities of synthesis of medical images using neural diffusion models. First, we use a pre-trained DALLE2 model to generate lungs X-Ray and CT images from an input text prompt. Second, we train a stable diffusion model with 3165 X-Ray images and generate synthetic images. We evaluate the synthetic image data through a qualitative analysis where two independent radiologists label randomly chosen samples from the generated data as real, fake, or unsure. 
Results demonstrate that images generated with the diffusion model can translate characteristics that are otherwise very specific to certain medical conditions in chest X-Ray or CT images. Careful tuning of the model can be very promising. 
To the best of our knowledge, this is the first attempt to generate lungs X-Ray and CT images using neural diffusion models. This work aims to introduce a new dimension in artificial intelligence for medical imaging. Given that this is a new topic, the paper will serve as an introduction and motivation for the research community to explore the potential of diffusion models for medical image synthesis. We have released the synthetic images on https://www.kaggle.com/datasets/hazrat/awesomelungs. 
\keywords{Diffusion models \and Generative models \and Artificial Intelligence \and Medical imaging \and lungs \and CT \and X-Ray}
\end{abstract}
\section{Introduction}
During the last decade, there has been a surge in studies on generative models for medical image synthesis \cite{xin2019gansreview,jiang2020covid}. Generative Adversarial Networks (GANs) and deep autoencoders are two primary examples of deep generative models that have shown remarkable advancements in synthesis, denoising, and super-resolution of medical images \cite{xin2019gansreview,chen2017deepautoencoder}. Many studies have shown the great potential of GANs to generate realistic magnetic resonance imaging (MRI), Computed Tomography (CT), or X-Ray images that can help in training artificial intelligence (AI) models \cite{xin2019gansreview,AA11,AA3,munawar2020segmentation}. 
\begin{figure}[thb!]
    \centering
    \includegraphics[width=\linewidth]{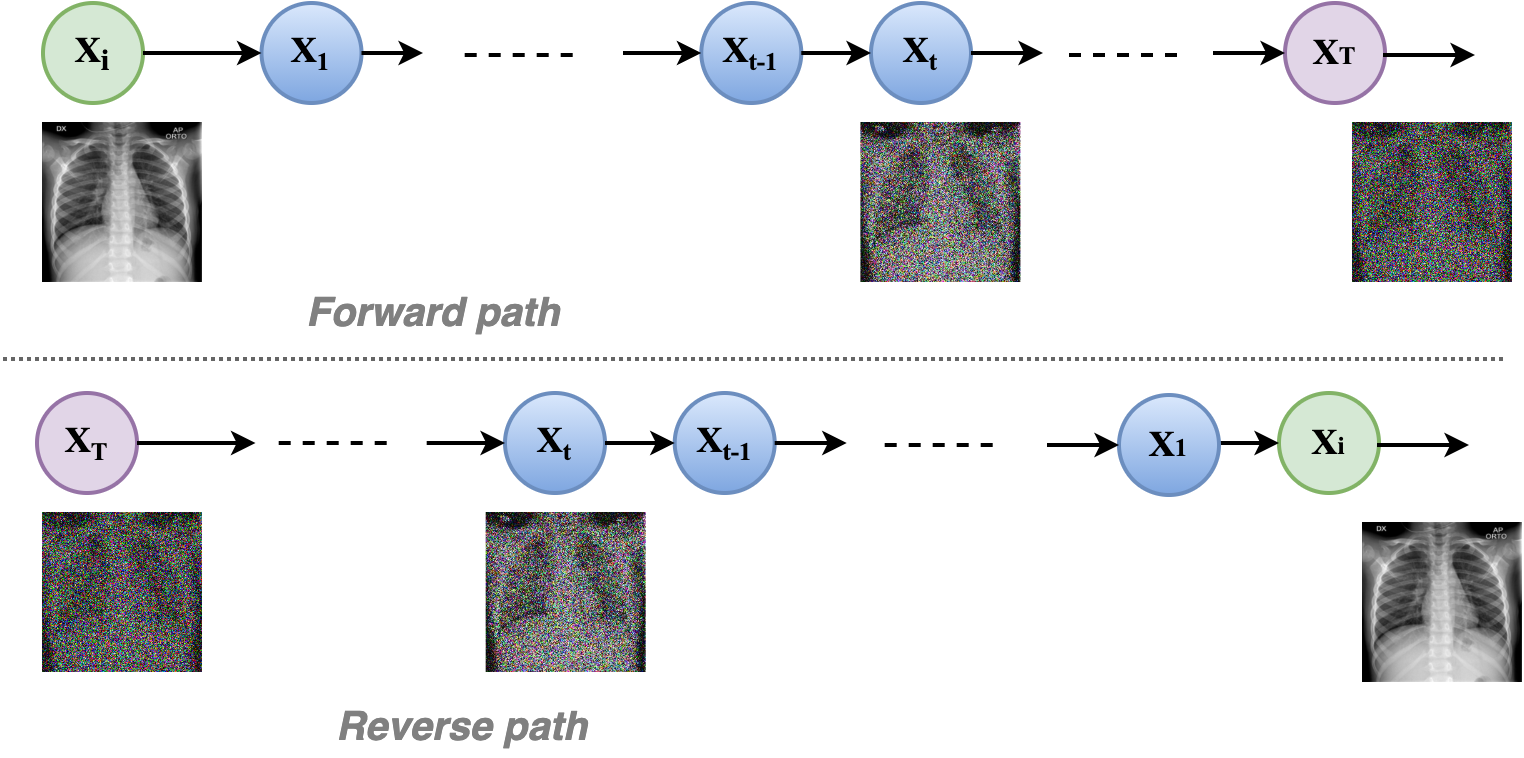}
    \caption{Forward pass and reverse pass in diffusion model training. Figure modified from \cite{AA10}.}
    \label{fig:forwardpass}
\end{figure}
With the recent success of neural diffusion models for the synthesis of natural images \cite{dhariwal2021,aditya2022clip}, there is now an increasing interest in exploring the potential of neural diffusion models to generate medical images. For generating natural images such as art images, objects, models such as DALLE2\footnote{https://openai.com/}, Mid-Journey\footnote{https://www.midjourney.com/home/}, and Stable Diffusion\footnote{https://github.com/CompVis/stable-diffusion} have pushed the state-of-art. Amongst the three, only the latter is available with open-source code. Compared to GANs, diffusion models are becoming popular for their training stability. 

A diffusion model, in simple words, is a parameterized Markov chain trained using variational inference. The transition is learned through a diffusion that adds noise to the data. In principle, the diffusion model transforms the input data into noisy data by adding Gaussian noise and then recovers the data distribution by reversing the noise. Once the model learns the distribution, it can generate useful data from random noise input. So, diffusion models transform a latent encoded representation into a more meaningful representation of image data. In this context, diffusion models can be compared to denoising autoencoders. As shown in Figure \ref{fig:forwardpass}, the overall process can be summarized as a two-step phenomenon, the forward pass, i.e., the transformation of the data distribution to noise ($X_i$ to $X_T$), and the reverse pass, i.e., reversing the noise distribution to data distribution ($X_T$ to $X_i$). Training a diffusion model implies the learning of the reversing process i.e., $p(x_{t-1} | x_t)$. The diffusion model can be implemented by using a neural network for the forward and reverse training steps. However, the architecture must have the same input and output dimensions.

While previously, the generating ability of diffusion models was mostly used for unconditional generation of data, more recent attempts have shown conditioned generation by introducing guided-diffusion models \cite{aditya2022clip,AA5,AA7}. These works have demonstrated the generation of photo-realistic images guided by the context of the input text or image. The existing use cases of diffusion models comprise text-to-image applications, i.e., generating images according to a given text prompt. In addition, Han et al., \cite{AA8} presented a classification and regression diffusion model (CARD), and demonstrated the use of the diffusion model for classification as well as regression tasks. In CARD, the authors approached the task of supervised learning using generative modeling conditioned on the class labels. Though the objective was not to claim state-of-the-art results, the method has shown promising results on the benchmark dataset. For CIFAR-10 classification, the model reached an accuracy of 90.9\%. 

Given the potential of diffusion models to learn the representation, one can expect their potential to generate a diverse set of medical images. Furthermore, they can add a new dimension to existing approaches for medical image applications, such as noise adaptation, noise removal, super-resolution, domain-to-domain translation, and data augmentation. 
To the best of our knowledge, no work other than the recent pre-print \cite{AA4}, exists currently on the synthesis of medical images using neural diffusion models. Walter et al, \cite{AA4} used latent diffusion models to generate T1w MRI images of the brain. Using 31,740 brain MRI images from the UK Biobank, they have generated a stack of 100,000 images conditioned on key variables such as age, sex, and brain volume. In this work, we explore neural diffusion models to generate synthetic images of lung CT and X-Ray. We use the DALLE2 model and the stable diffusion model to generate the images and present them to two radiologists for their feedback. We then summarize the feedback received from the radiologists and identify some of the challenges in using the neural diffusion model for medical image synthesis. 

The remaining paper is organized as Section \ref{sec:method} explains the methodology of our work. Section~\ref{sec:results} presents the results of generating lung CT and X-Ray images, while Section \ref{sec:discussion} provides insights into the results and also highlights the limitations of the approach. Finally, Section~\ref{sec:conclusion} concludes the paper. 

\section{Methodology}
\label{sec:method}
In this work, we devised two experiments for generating synthetic images of lungs X-Ray and CT. In the first experiment, we used the \emph{OpenAI} DALLE2 API \footnote{https://openai.com/} to generate images based on the input text. The DALLE2 model recently gained much attention for its ability to generate photo-realistic images of objects given a certain input text. Using the API, we generated multiple images of lungs CT and X-Ray. We then presented a randomly selected set of the generated images to two trained radiologists. We asked the radiologists for two key tasks. First, we asked them to label each image as real, fake, or uncertain about, as per their perceived understanding. Second, we asked them to provide a brief description of the possible information related to lung condition or diagnosis of disease (for example, normal lungs, severely damaged lungs, pneumonia-affected lungs, etc). The radiologists did not have prior information on the labels of the images. In fact, all the images that we presented to the radiologists were synthetic. The radiologists did not know each other and performed the tasks independently. Of the two radiologists, one radiologist had prior knowledge of artificial intelligence and generative models, while the other radiologist was naïve to deep generative models. 

In the second experiment, we used the stable diffusion model \cite{AA15}. We trained the stable diffusion model using 3165 X-Ray images from \cite{AA14}. We resized the images to 256 by 256 resolution. No other pre-processing was done. Using the X-Ray images, we trained a stable diffusion model on a server equipped with NVIDIA Quadro RTX 8000 GPU with a 48 GB memory. We set the batch size equal to 32 and ran the training for 700000 training steps. 

\section{Results}
\label{sec:results}
\begin{figure}[!htb]
     \centering
     \begin{subfigure}[t]{0.22\textwidth}
         \centering
         \includegraphics[width=\textwidth]{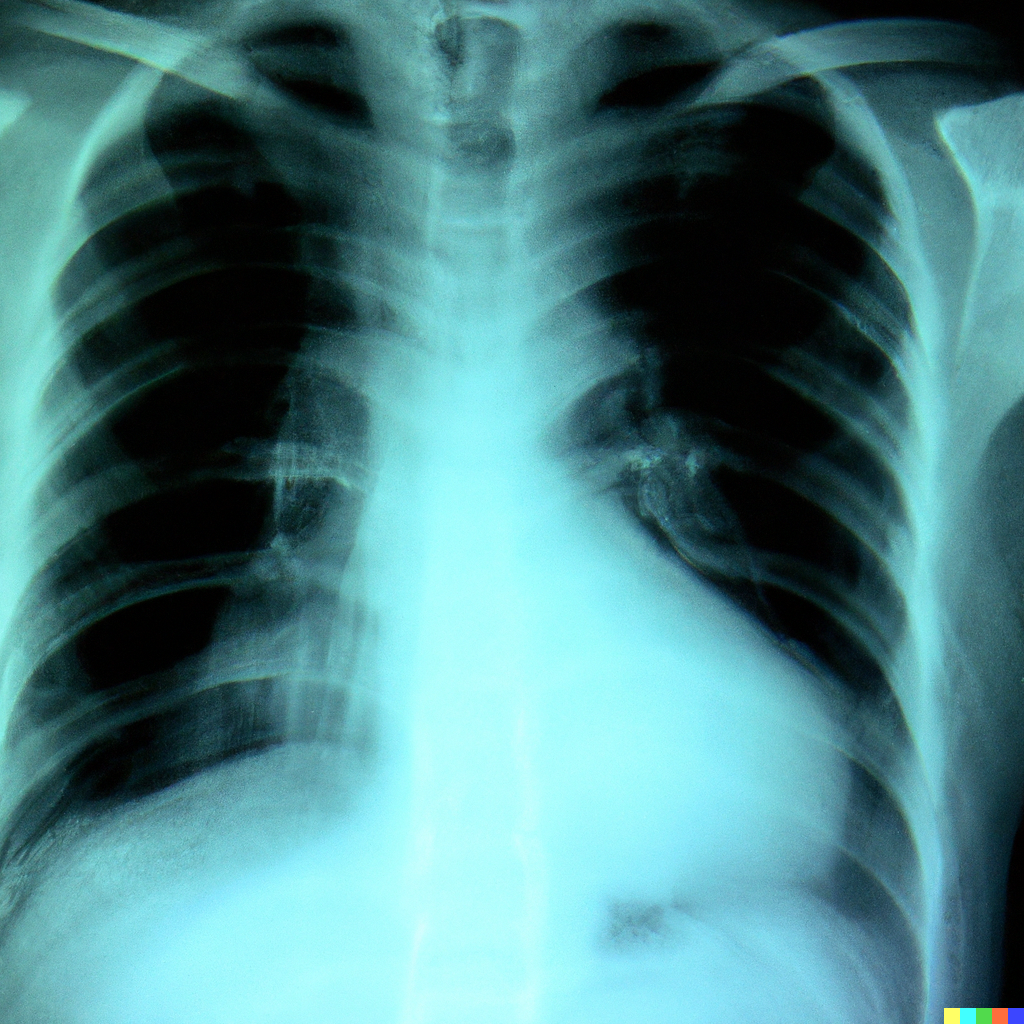}
         \label{FFHQB}
     \end{subfigure}
     \begin{subfigure}[t]{0.22\textwidth}
         \centering
         \includegraphics[width=\textwidth]{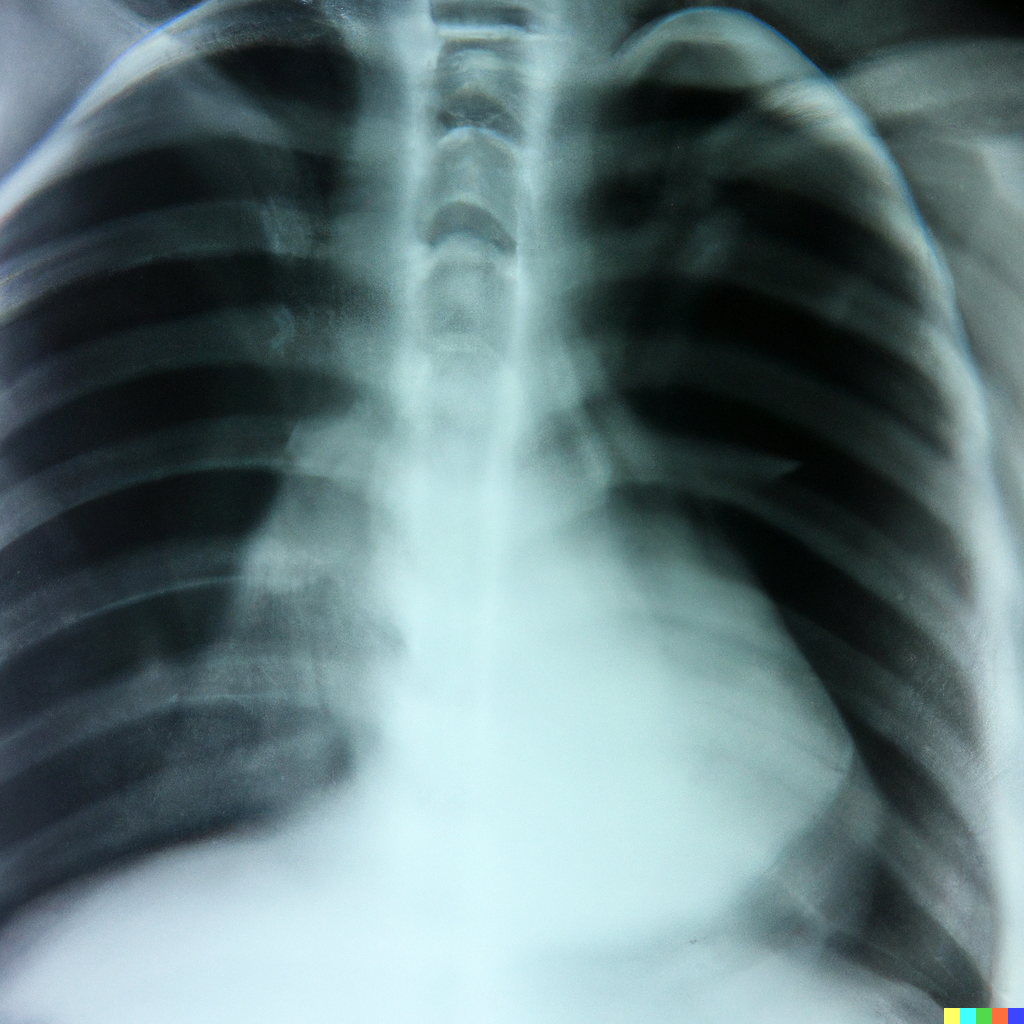}
         \label{AFHQB}
     \end{subfigure}
     \begin{subfigure}[t]{0.22\textwidth}
         \centering
         \includegraphics[width=\textwidth]{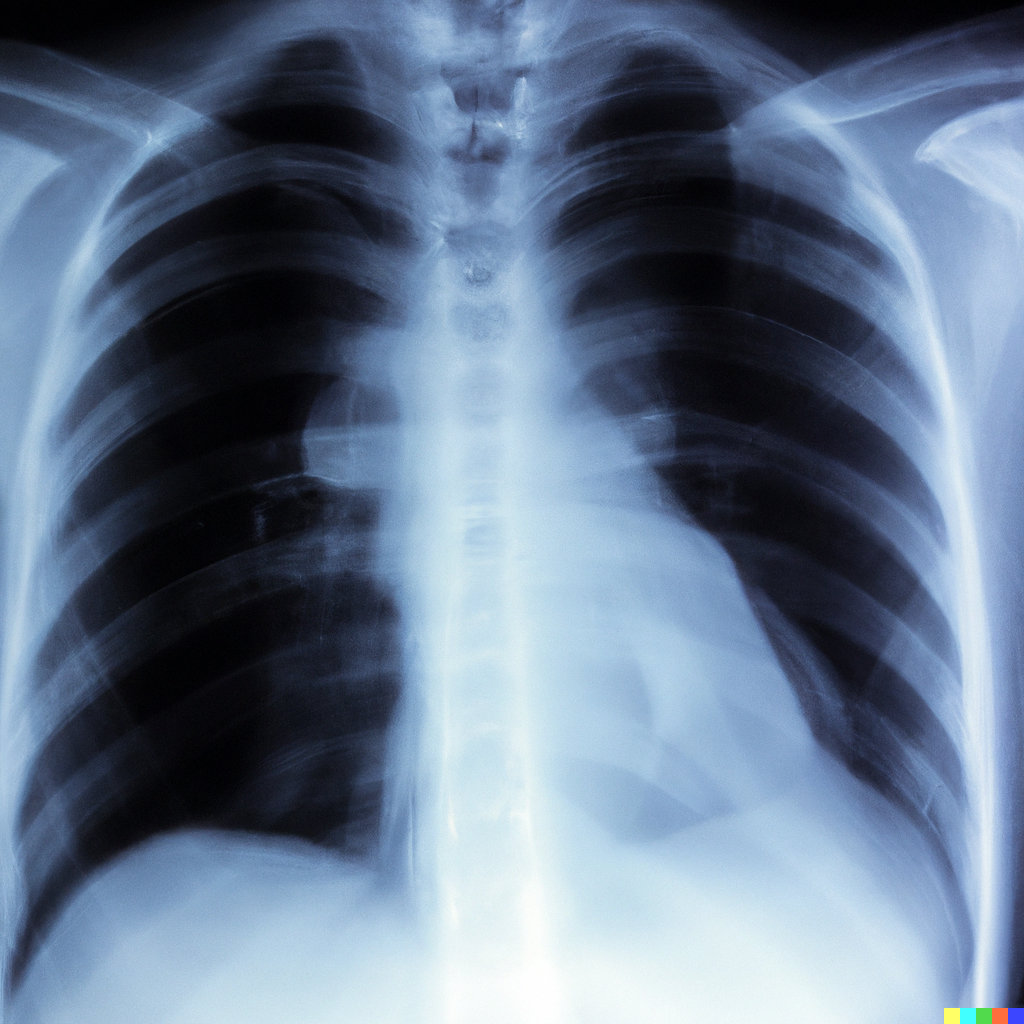}
         \label{AFHQWB}
     \end{subfigure}
     \begin{subfigure}[t]{0.22\textwidth}
         \centering
         \includegraphics[width=\textwidth]{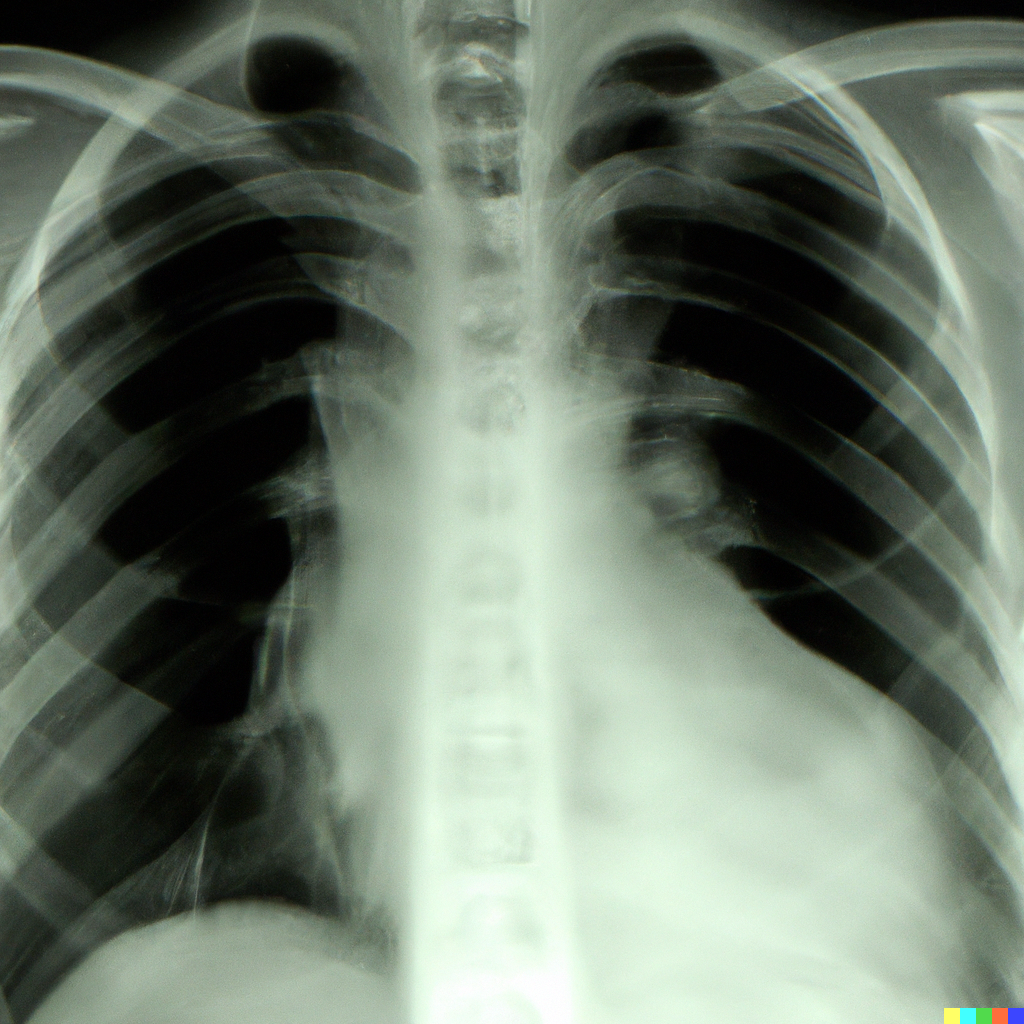}
         \label{fig:five over x}
     \end{subfigure}
        \caption{Samples of lungs X-Ray images generated with the diffusion model.}
        \label{fig:samplexray}
\end{figure}

\begin{figure}[!htb]
     \centering
     \begin{subfigure}[t]{0.22\textwidth}
         \centering
         \includegraphics[width=\textwidth]{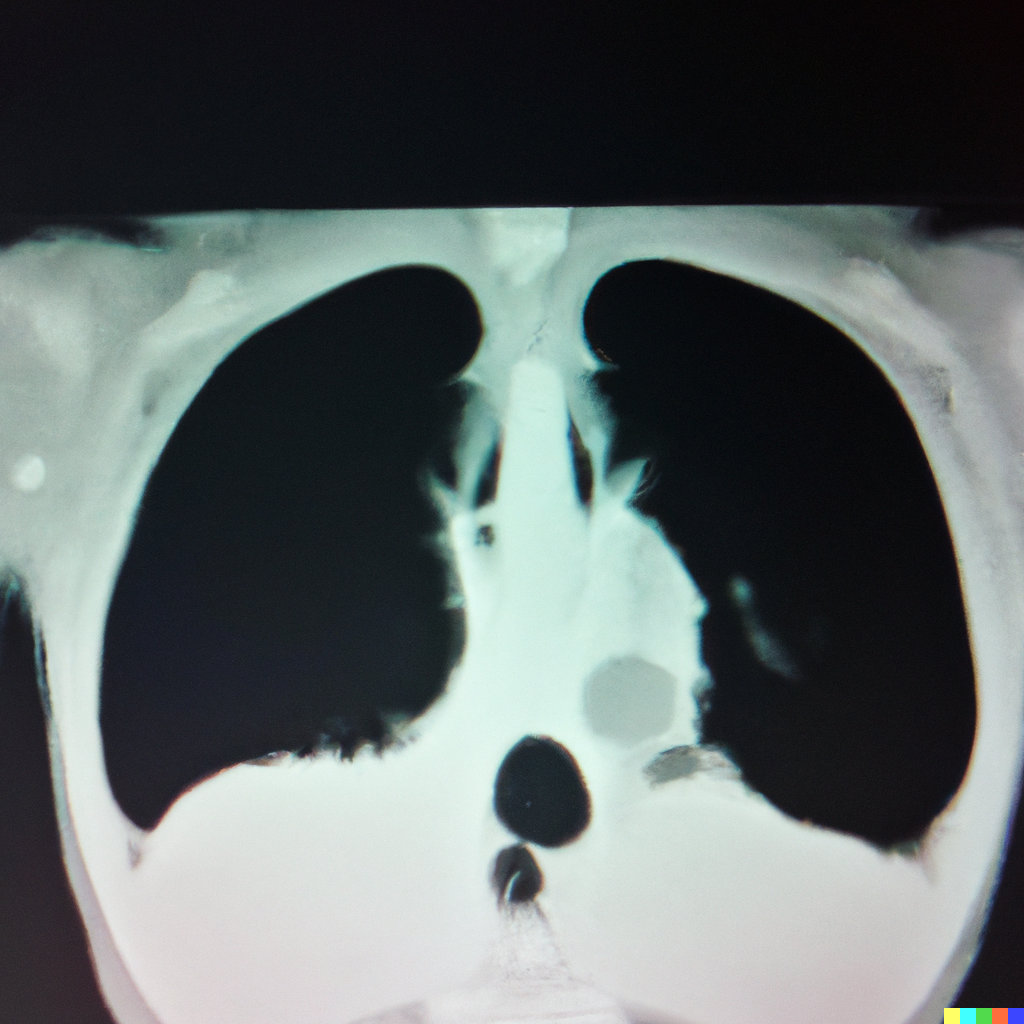}
         \label{FFHQB}
     \end{subfigure}
     \begin{subfigure}[t]{0.22\textwidth}
         \centering
         \includegraphics[width=\textwidth]{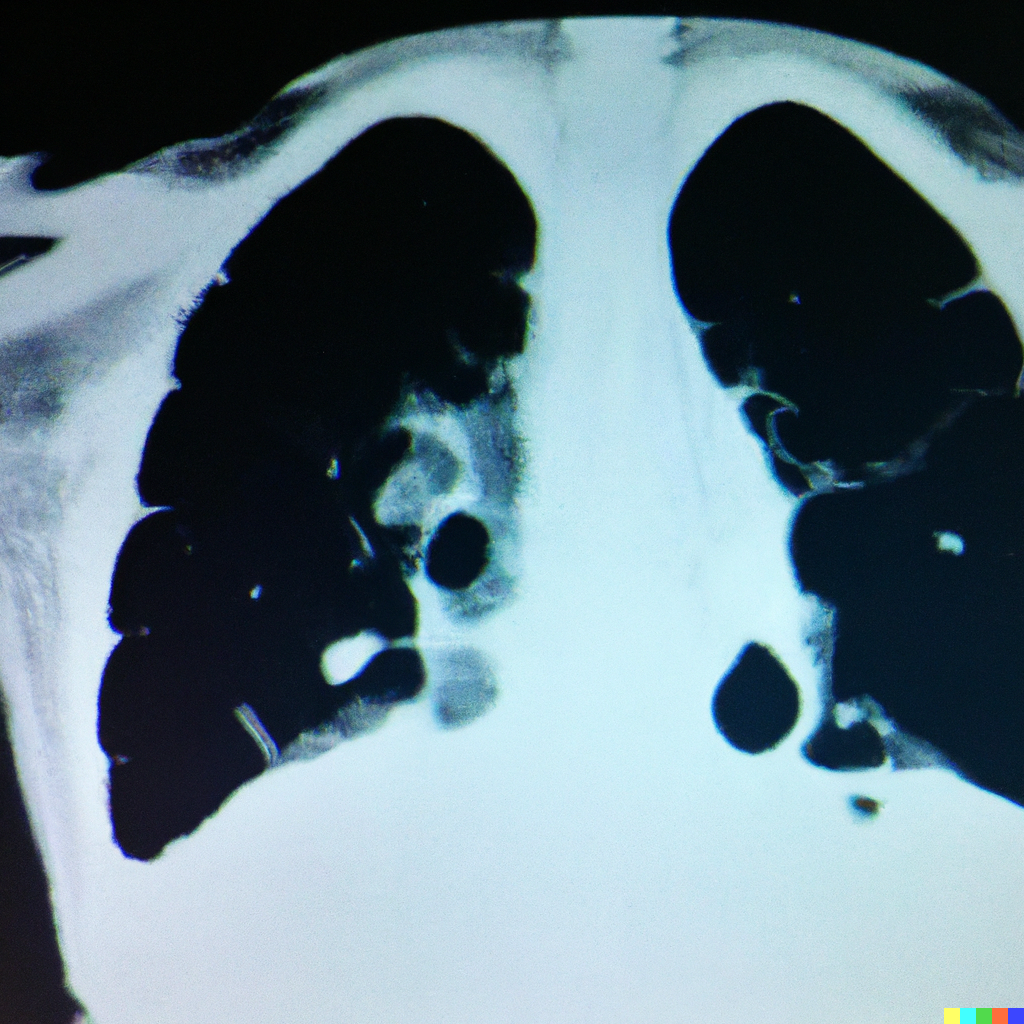}
         \label{AFHQB}
     \end{subfigure}
     \begin{subfigure}[t]{0.22\textwidth}
         \centering
         \includegraphics[width=\textwidth]{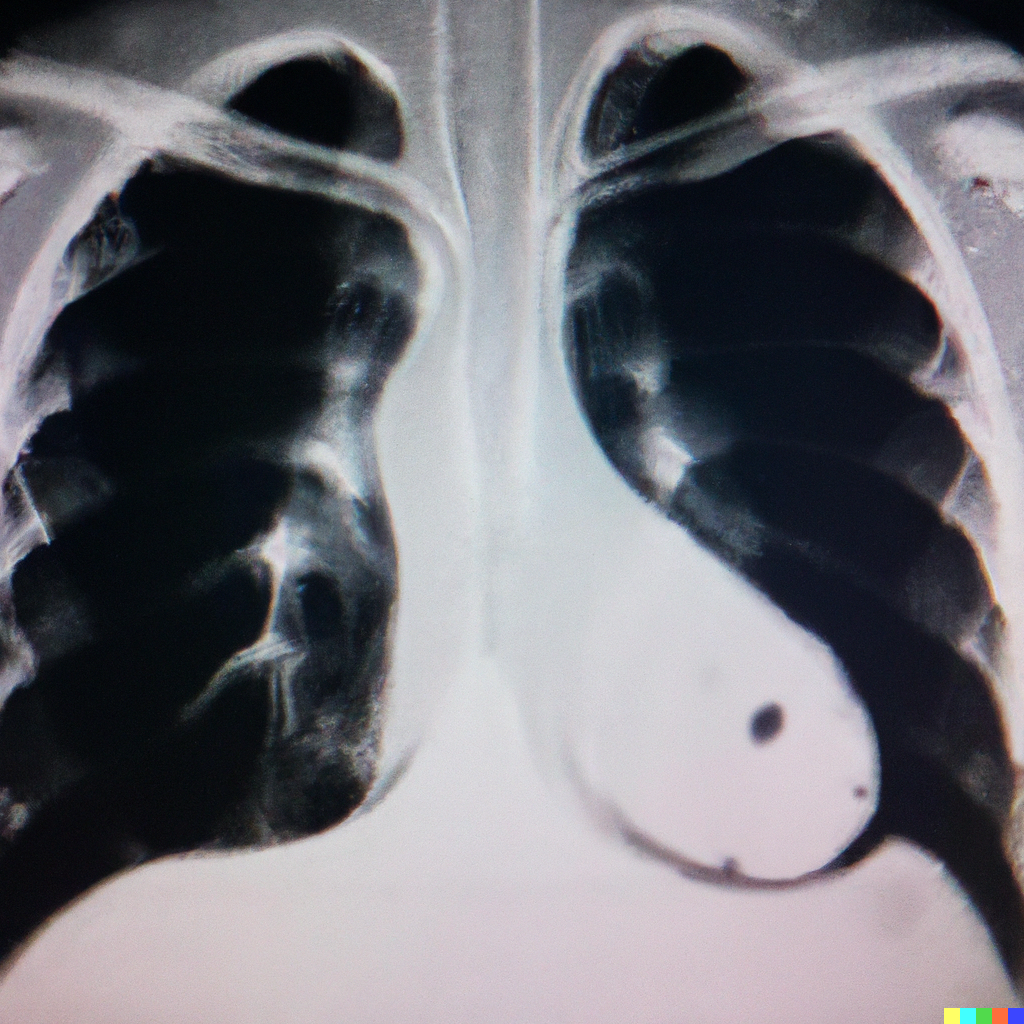}
         \label{AFHQWB}
     \end{subfigure}
     \begin{subfigure}[t]{0.22\textwidth}
         \centering
         \includegraphics[width=\textwidth]{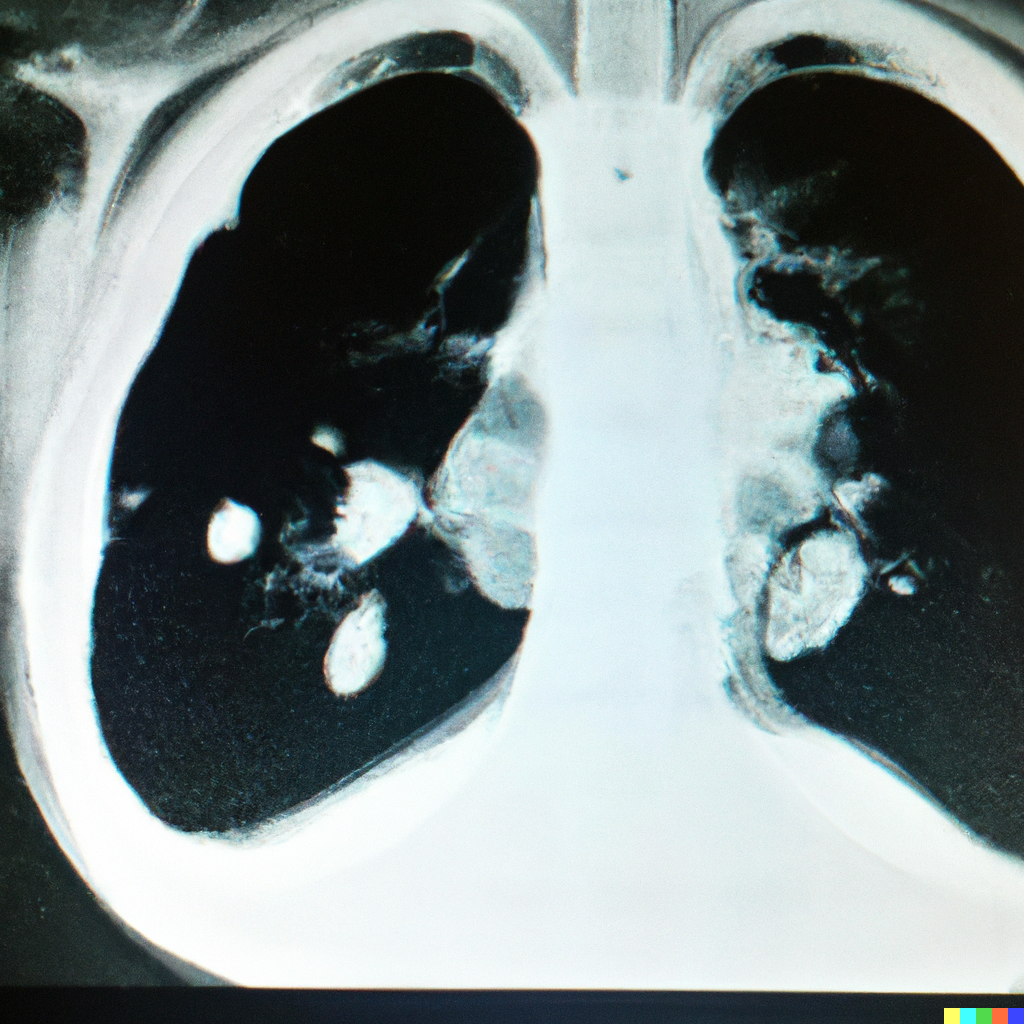}
         \label{fig:five over x}
     \end{subfigure}
        \caption{Samples of lungs CT images generated with diffusion model.}
        \label{fig:samplect}
\end{figure}

Using the DALLE2 API, we generated a total of 150 images. We have uploaded the synthetic images to Kaggle \footnote{https://www.kaggle.com/datasets/hazrat/awesomelungs}. We believe the number of generated images is only limited by the tokens available to us. Sample X-Ray and CT images are shown in Figure \ref{fig:samplexray} and Figure \ref{fig:samplect}, respectively. Out of 40 images that we presented to the radiologists, radiologist $\mathcal{A}$ identified 14 X-Ray images and three CT images as real, while 4 X-Ray and 17 CT images as fake. Radiologist $\mathcal{A}$ labeled two X-Ray images as unsure. The second radiologist (radiologist $\mathcal{B}$) identified 10 X-Ray images and only two CT images as real, while all the remaining images as fake. 

Agreement between radiologists: Of the 20 CT images, only three images were labeled as real by both radiologists. Similarly, five X-Ray images were marked as real by both radiologists. There were 2 X-Ray and 2 CT images for which both the radiologists were uncertain. 

For task 2, where we asked the radiologist to provide a description of what the images may reveal, the radiologists made some interesting observations. For example, some descriptions are listed in Table \ref{tab:remarks}. These descriptions clearly reveal that some of the images carried representations similar to real X-Ray or CT images, and the model was able to generate features that are specific lung conditions. 

\begin{table}[!htb]
\begin{center}
\caption{Samples of remarks from radiologists (no-specific order)}
\label{tab:remarks}
\begin{tabular}{p{3cm}|p{6cm}}
\hline
\textbf{Image modality} & \textbf{Remarks*}\\
\hline
CT      & Possible effusions\\
        & Pneumonia \\
\hline
X-Ray   & Left lower lobe effusions \\
        & Possibility of pneumonia \\
        & Bilateral infection \\
\hline
\multicolumn{2}{l}{\small *The remarks do not imply a definite decision.} \\
\end{tabular}
\end{center}
\vspace{-8mm}
\end{table}


\vspace{-2mm}


\begin{figure}[!htb]
     \centering
     \begin{subfigure}[t]{0.22\textwidth}
         \centering
         \includegraphics[width=\textwidth]{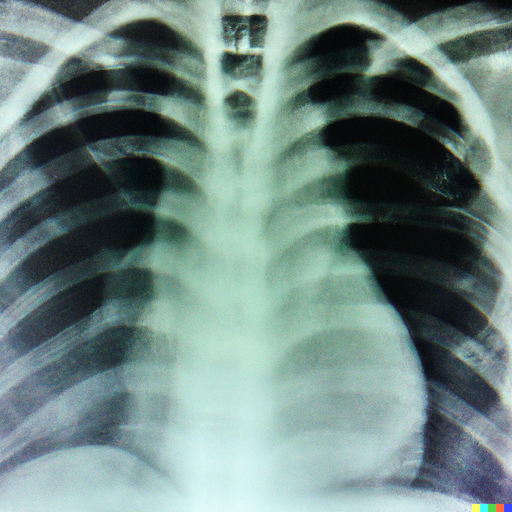}
         \label{fig:im27}
     \end{subfigure}
     \begin{subfigure}[t]{0.22\textwidth}
         \centering
         \includegraphics[width=\textwidth]{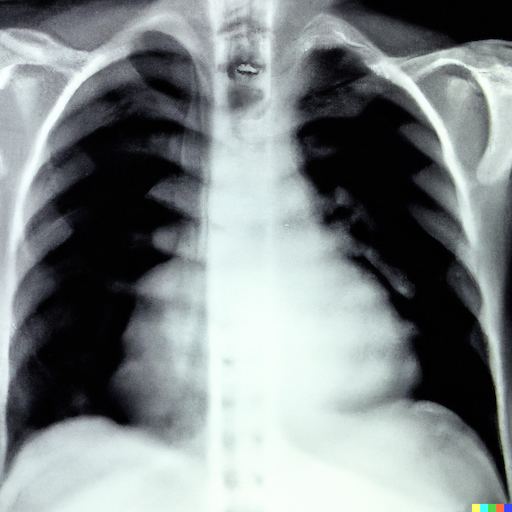}
         \label{fig:im36}
     \end{subfigure}
    \begin{subfigure}[t]{0.22\textwidth}
         \centering
         \includegraphics[width=\textwidth]{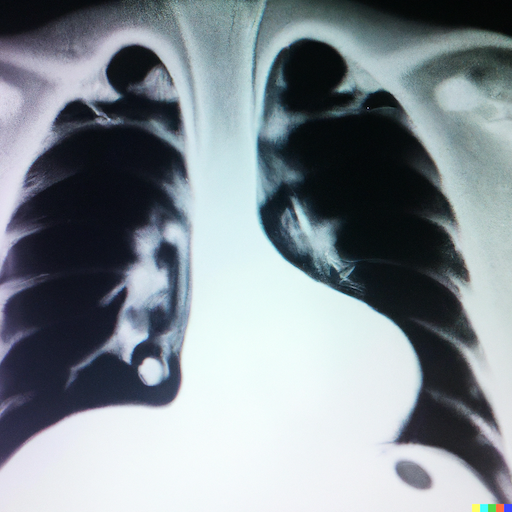}
         \label{AFHQWB}
     \end{subfigure}
     \begin{subfigure}[t]{0.22\textwidth}
         \centering
         \includegraphics[width=\textwidth]{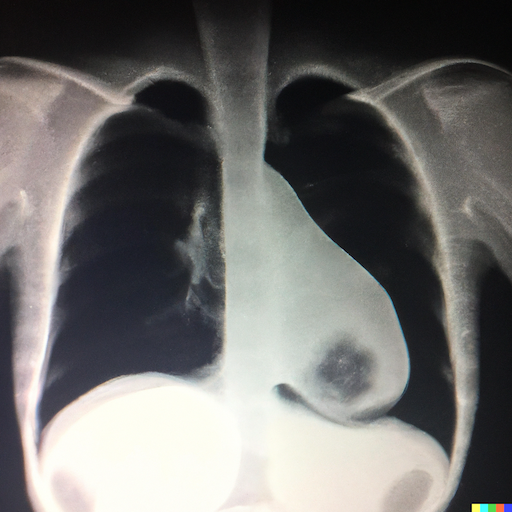}
         \label{fig:five over x}
     \end{subfigure}
    \caption{Samples of synthetic images for lungs X-Ray (left two images) and CT (right two images) identified as fake by at least two radiologists.}
    \label{fig:fakeimages}
\end{figure}
\begin{figure}[thb!]
    \centering
    \includegraphics[width=\linewidth]{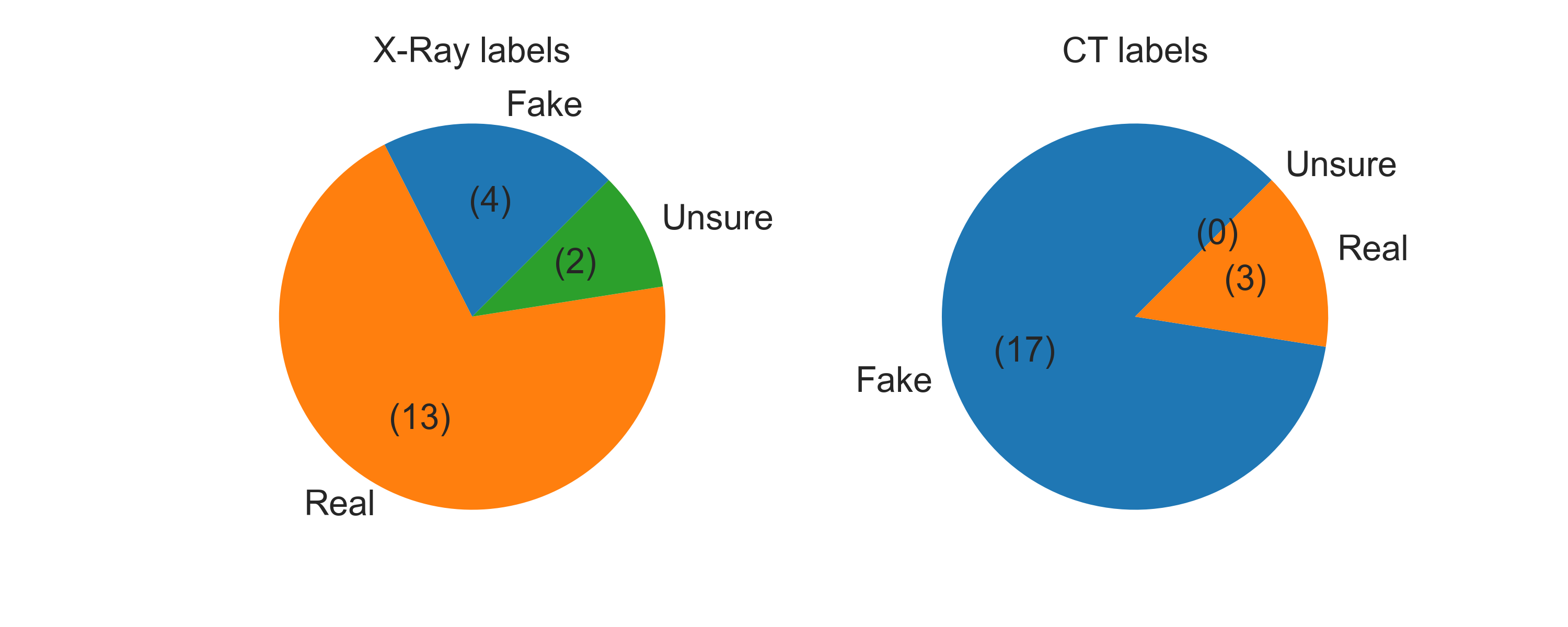}
    \caption{Evaluation by Radiologist A}
    \label{fig:forwardpass}
\end{figure}
\section{Discussion}
\label{sec:discussion}
Some of the generated images lacked the characteristics of realistic images and were quickly identified by the radiologists as fake. These images were termed as having unusual ribs appearance or showing unusual exposure. Similarly, it was easy to spot big vessels contour and lung fields that appeared to have been drawn and not imaged. One key observation for fake images was that the trachea is visible behind the heart shadow, which does not happen in real X-Ray imaging. 
A few sample images that were termed fake by at least two radiologists are shown in Figure \ref{fig:fakeimages}.  
Many of the generated images from the pre-trained model clearly lacked the characteristics of realistic images and were quickly identified by the radiologists as fake. These images were termed as having unusual ribs appearance, strange clavicle appearance, or showing unusual exposure. 

\subsection{Limitations}
One challenge identified in diffusion models is the limited ability to produce details in complex scenes \cite{aditya2022clip}. So, generating complex medical images would need to be complemented with noise adaptation or super-resolution techniques \cite{AA3}. 
Like many other AI models, diffusion model training is prone to bias in the dataset; for example, unbalanced representation of medical conditions in the input X-Ray or CT image or inherent noise in data. Thus, the synthetic data from such a diffusion model will also carry the bias. Eventually, if the generated data are made public and used for onward model training, the bias may turn into a cascade behavior and will be further augmented \cite{AA9}. 
The model has been used pretty much as a black-box model; hence, not much explainability can be offered on how certain images were generated. Unlike the work reported in Walter et al., \cite{AA4}, our generated images are not conditioned on additional variables such as gender, age, etc. Diffusion models are very slow to train as they require the number of training steps to be in the order of several hundred thousand. Our training took around one day for 100k training steps. 

\section{Conclusion and Future Work}
\label{sec:conclusion}
In this work, we have demonstrated the potential of neural diffusion models for the synthesis of lungs X-Ray and CT images. Though the radiologists spotted many images as fake, few images were still labeled as real by them. The labeling from the radiologists reflects that some of the generated X-Ray images carried a great resemblance to real images. However, the identification of fake images was straightforward for the CT images. Through qualitative analysis of the generated images, we showed that neural diffusion models have great potential to learn complex representations of medical images. Although the performance of diffusion models is superior to GANs-based methods for synthesizing natural images, research efforts on the diffusion model for medical image synthesis have yet to mature. 

\section{Acknowledgments}
The authors are grateful to Surendra Maharjan from Indiana University Purdue University  Indianapolis, USA, for useful comments on this work. The authors are thankful to Dr. Jens Schneider for facilitating the GPU access.  

%
%
%

\bibliographystyle{ieeetr}
\bibliography{paper.bib}





\end{document}